\documentclass[12pt,twocolumn]{iopart}
\usepackage{iopams}
\usepackage{graphicx}
\usepackage[usenames]{color}
\usepackage[normalem]{ulem}
\usepackage{soul}
\newcommand{\cpc}{\emph{Comp. Phys. Commun.\ }}

\begin{document}

\title[Collisionally inhomogeneous BEC in a bichromatic optical lattice]{Collisionally inhomogeneous Bose-Einstein condensates with binary and three body interactions in a bichromatic optical lattice}

\author{J. B. Sudharsan$^{1}$, R. Radha$^{1}$ and P. Muruganandam$^{2}$}
\address{$^{1}$Center for Nonlinear Science, PG and Research Department of Physics, Government College for Women (Autonomous), Kumbakonam 612001, Tamil Nadu, India\\ $^{2}$School of Physics, Bharathidasan University, Palkalaiperur Campus, Tiruchirappalli 620024, Tamilnadu, India}

\ead{radha\_ramaswamy@yahoo.com$ ^1$}
\ead{anand@cnld.bdu.ac.in$ ^2$}

\begin{abstract}
We study the impact of collisionally inhomogeneous binary and three body interaction on Bose-Einstein condensates (BECs) of a dilute gas in a bichromatic optical lattice. We observe that the localized matter wave density which decreases after the introduction of repulsive spatially inhomogeneous binary interaction can be sustained by the addition of constant attractive binary strength in equal amounts. If the balance between repulsive spatially inhomogeneous binary interaction and constant attractive interaction is disturbed, the condensates collapse. Reversal of sign of interaction ensures the longevity of BECs. Any imbalance between attractive spatially inhomogeneous interaction and constant repulsive interaction either results in the collapse of BECs or in the occupation of the condensates at multiple sites on either sides. The introduction of a weak three body interaction in phase with the binary interaction increases the extent of instability of BECs. Reversing the sign of spatially inhomogeneous and constant interaction enhances the stability of BECs.
\end{abstract}

\maketitle

\section{Introduction}

The experimental realization of Bose-Einstein condensation (BEC) in dilute bosonic gases \cite{Anderson1995} has completely revolutionized the understanding of ultra cold matter and has resulted in an explosion of interest both from the experimental and theoretical point of view in the domain of ultra cold atoms/molecules. These investigations have been carried out primarily by the mean field description of a BEC governed by the Gross-Pitaevskii (GP) equation which is a variable coefficient nonlinear Schrodinger (NLS) equation of the following form \cite{Gross1961}
\begin{equation}
\fl i\hbar\frac{\partial\Psi({\bf r}, \tau)}{\partial\tau}=\left[-\frac{\hbar^2}{2m} \nabla^2 +v({\bf r})+g({\bf r})\vert\Psi({\bf r}, \tau)\vert^2\right]\Psi({\bf r}, \tau).
\end{equation}
In the above equation, $v({\bf r})$ represents the external trapping potential and $g({\bf r})$ the binary interatomic interaction. However, it was realized that the mean field description of a BEC with a binary interaction alone was inadequate for generating high density BECs \cite{Tomio2000}. In this context, the impact of a three body interaction \cite{Abdullaev2001} in generating high density condensates assumes greater significance. This means that a more accurate mean field description should take into account both binary and three body interaction. In this connection, the dynamics of BECs with both binary and three body interactions was analysed \cite{Tomio2000,Abdullaev2001,Kohler2002,Li2005,Kengne2008,RameshKumar2010,Sabari2010} and it was found that the high density condensates were found to be stable for attractive binary and repulsive three body interactions. The interplay between cubic and quintic nonlinearties in stabilizing vortex solitons had also been investigated \cite{Mihalache2002}. Eventhough the scattering length in general could vary as a function of space and time, the investigation of BECs are mainly centred around the temporal variation of interaction strengths \cite{ZXLiang2005}. It was realized that the spatial variation of the laser field intensity \cite{Fedichev1996} by proper choice of the resonance detuning can lead to the spatial dependence of the atomic scattering lengths leading to the so-called \emph{collisionally inhomogeneous BECs}. Such spatial dependence of the scattering lengths which can be implemented utilizing a spatially inhomogeneous external magnetic field in the vicinity of a Feshbach resonance \cite{Xiong2005} renders the collisional dynamics inhomogeneous across the BECs. The resulting so-called \emph{collisionally inhomogeneous environment} provides a variety of interesting and previously unexplored dynamical phenomena and potential applications like adiabatic compression of matter waves \cite{Theocharis2005,Rodas2005,Theocharis2006}, atomic soliton emission and atom lasers \cite{Garnier2006}, enhancement of the transmittivity of matter waves through barriers \cite{Kartashow2011}, dynamical trapping of matter waves solutions \cite{Siva2006} etc. The investigation of collisionally inhomogeneous interactions with linear and nonlinear  lattices \cite{Kartashov2008} and the recent identification of gap solitons in periodic media supported by localized nonlinearites \cite{dror2011} underscore the impact of spatially inhomogeneous environment on optical solitons and BECs. The impact of spatially inhomogeneous interaction on the condensates in a monochromatic optical lattice \cite{Golam2012} has been explored recently. This paper is a nascent attempt to study the combined impact of collisionally inhomogeneous binary and collisionally inhomogeneous three body interactions on the condensates of a dilute gas in a bichormatic optic lattice potential.

The organization of this paper is as follows. In sec. 2, we present the mathematical model for the investigation of the impact of spatially inhomogeneous two and three body interactions on BECs. The variational solution of the time-independent GP equation is also given. In sec. 3,  we numerically solve the GP equation and discuss the impact of collisionally inhomogeneous binary and three body interactions. In sec. 4,  we present a brief discussion and concluding remarks.

\section{ Model and evolution equation}
Considering both binary and three body interactions, the BEC in three dimensions is described by the following mean field GP equation
\begin{equation}
 \fl i\hbar\frac{\partial\Psi({\bf r}, \tau)}{\partial\tau}=\left[-\frac{\hbar^2\nabla^2}{2m}+U({\bf r})+g({\bf r})\vert\Psi({\bf r}, \tau)\vert^2+k({\bf r})\vert\Psi({\bf r}, \tau)\vert^4\right]\Psi({\bf r}, \tau), \label{eq:gpe3d}
\end{equation}
where the normalization is given by $\int_{-\infty}^\infty \vert \Psi({\bf r}, \tau)\vert^2 dr= N$, $\tau$ the time, $N$ the number of atoms, $U({\bf r})$ the bi-chromatic lattice, $g({\bf r})$ and $k({\bf r})$ are the space-dependent two and three body interactions respectively.
\paragraph{•}
The bi-chromatic optic lattice potential is generated by two standing wave polarized laser beams of incommensurate wavelengths whose generic form is given by [14]:
%\begin{numparts}
\begin{eqnarray}
U({\bf r}) = \sum^2_{i=1} s_iE_i\sin^2(k_i.{\bf r}), \label{eq:pot:sin} %\\
%U({\bf r}) = \sum^2_{i=1} s_iE_i\cos^2(k_i.{\bf r}),  \label{eq:pot:cos}
\end{eqnarray}\color{black}%
%\end{numparts}%
where $s_i$, $i=1, 2$ are the amplitudes of the OL potentials in units of respective recoil energies $E=2\pi^2\hbar^2 / m\lambda'^2_i$,  $k_i= 2\pi / \lambda'_i$ are the respective wave numbers and $\lambda'_i$ are the wavelengths. In the actual experiments of Roati et al \cite{Roati2008}, the wavelengths of bi-chormatic OL wavelengths are $\lambda'_1= 1032 nm$ and $\lambda'_2= 862nm$.
\paragraph{•}
For a cigar-shaped trap with strong transverse confinement, it is appropriate to consider a 1D reduction of equation (\ref{eq:gpe3d}) by freezing the transverse dynamics to the respective ground state and integrating over the transverse variables $y$ and $z$. The resulting quasi-1D dimensionless GP equation for cigar-shaped BECs is given by,
\begin{equation}
\fl i\frac{\partial\phi(x,t)}{\partial t} = \left[-\frac{1}{2}\frac{\partial^{2}}{\partial
x^2}+V(x)+g(x)\vert\phi(x, t)\vert^{2}+k(x)\vert\phi(x, t)\vert^{4}\right]\phi(x, t),\label{eq:gpe}
\end{equation}
where the normalization is $\int_{-\infty}^\infty \vert \phi(x, t)\vert^2 dx= N$, length is in units of $\sqrt{\hbar / m\omega_\perp}$, time is in $\omega^{-1}_\perp$ and energy is in units of $\hbar\omega_\perp$.  The potential in equation (\ref{eq:pot:sin}) is now defined by one of the following expressions
%\begin{numparts}
\begin{eqnarray}
V(x) & = & \sum_{i=1}^2 \frac{4\pi^2s_i}{\lambda^2_i} \sin^2 \left(\frac{2\pi}{\lambda_i} x \right), \label{pot:sin}
%\\ V(x) & = & \sum_{i=1}^2 \frac{4\pi^2s_i}{\lambda^2_i} \cos^2 \left(\frac{2\pi}{\lambda_i} x \right).\label{pot:cos}
\end{eqnarray}\color{black}%
%\end{numparts}%
$g(x)=2a_x/a_\perp$ and $k(x)$ are the strengths of space-dependent binary and three body interactions respectively. $a_x$ is the space dependent s-wave scattering length which can be tuned to any desire value using the Feshbach resonance technique.

The dependence of $a_x$ on the spatial co-ordinate $x$ is given by \cite{Fedichev1996,Theis2004,Sakaguchi2005}
\begin{equation}
a_{x} = a_0+\frac{\alpha I(x)}{\delta+\beta I(x)},
\end{equation}
where $I(x)$ is the intensity of laser light and $a_0$ stands for the scattering length in the absence of light. The quantities $\alpha$ and $\beta$ are constants which depend on the detuning parameter $\delta$ of the laser. For a large detuning laser beam having Gaussian intensity variations, $g(x)$ and $k(x)$ can be written as \cite{Fedichev1996,Theis2004,Sakaguchi2005} ,
\begin{numparts}
\begin{eqnarray}
g(x)= \gamma_0+\gamma_1 \exp(-x^2/2), \\
k(x)= \eta_0+\eta_1 \exp(-x^2/2),
\end{eqnarray}
\end{numparts}%
where $\gamma_0$, $\eta_0$, $\gamma_1$ and $\eta_1$ are constants. The strength of the three body interaction $k(x)$ is always less than the two body interaction $g(x)$~\cite{Gammal2000}.
%as pointed out by Gammal et al.
For our present study, we consider that $k(x)$= 10\% of $g(x)$. Hence, the interaction $k(x)$ is dependent on the interaction $g(x)$ which means that one can control the three body interaction using Feshbach resonance by tuning $s$-wave scattering length \cite{Gammal2000}.

By eliminating the time dependence by $\phi(x)= \phi(x)\exp(-i\mu t)$, the stationary version of equation (\ref{eq:gpe3d}) is written as
\begin{equation}
\mu = \left[-\frac{1}{2}\frac{\partial^{2}}{\partial
x^2}+V(x)+g(x)\vert\phi(x)\vert^{2}+k(x)\vert\phi(x)\vert^{4}\right]\phi(x), \label{eq:gpei}
\end{equation}
$\mu$ is the chemical potential. %For writing equation (11), we have used $\phi(x)= \phi(x)e^{-i\mu t}$ with the normalization $\int_{-\infty}^{\infty} \vert\phi(x)\vert^2 dx= N$.
The Lagrangian for equation (\ref{eq:gpei}) is given by~\cite{Gammal2000}
\begin{equation}
L=\int{[\mu\vert\phi\vert^2-\frac{1}{2}\vert\nabla\phi\vert^2-V(x)\vert\phi\vert^2-\frac{g(x)}{2}\vert\phi\vert^4-\frac{k(x)}{3}\vert\phi\vert^6]dx-\mu}. \label{lag:dens}
\end{equation}
To apply the variational approximation, we assume a trial Gaussian function for $\phi(x)$ as
\begin{equation}
\phi(x)= \pi^{-\frac{1}{4}}\sqrt{\frac{\mathcal{N}}{w}}\exp\left(-\frac{x^2}{2w^2}\right)
\label{eq:trial}
\end{equation}
where $\mathcal{N}$ is the norm and $w$ is the width. Substituting equations (\ref{eq:trial}) and (\ref{pot:sin}) in equation (\ref{lag:dens}) and integrating overall space, we get the effective Lagrangian
  \begin{eqnarray}
\nonumber L_{\mathrm {eff}} = \mu{(\mathcal{N}-1)}+\frac{\mathcal{N}}{4w^2}+\mathcal{N}\sum_{i=1}^2\frac{A_i}{2}\left[1-\exp\left(-\alpha^2_iw^2\right)\right]+\frac{\gamma_0}{2\sqrt{2\pi}}\frac{\mathcal{N}^2}{w}\\
\hspace*{2cm}+\frac{\gamma_1}{\sqrt{2\pi}}\frac{\mathcal{N}^2}{w}\frac{1}{\sqrt{4+w^2}}+\frac{\eta_0}{3\pi\sqrt{3}}\frac{\mathcal{N}^3}{w^2}+\frac{\eta_1\sqrt{2}}{3\pi}\frac{\mathcal{N}^2}{w^2\sqrt{6+w^2}}.\label{eq:lag}
  \end{eqnarray}
where $A_i = 4\pi^2s_i/ \lambda_i^2, \alpha_i=2\pi/\lambda_i$. The first variational equation ${\partial L_{\mathrm {eff}}}/{\partial \mu}=0$ yields $\mathcal{N}=1$ which will be used in other variational equations. The second variational equation ${\partial L_{\mathrm {eff}}}/{\partial w}=0$ yields
 \begin{eqnarray}
\nonumber 2w^4\sum_{i=1}^2\left[A_i\alpha_i^2\exp\left(-\alpha^2_iw^2\right)\right] - \frac{\gamma_0 w}{\sqrt{2\pi}}-\sqrt{\frac{2}{\pi}}\frac{\gamma_1 w^3}{(w^2+4)^\frac{3}{2}}  - \sqrt{\frac{2}{\pi}}\frac{\gamma_1 w}{\sqrt{w^2+4}} \\
\hspace*{2cm} -\frac{4 \eta_0}{3 \pi \sqrt{3}}-\frac{\eta_1 2 \sqrt{2} w^2}{3 \pi} \frac{1}{(w^2+6)^\frac{3}{2}}  -\frac{4 \sqrt{2} \eta_1}{3 \pi}\frac{1}{\sqrt{w^2+6}}=1. \label{eq:vari}
 \end{eqnarray}
  and it determines the width $w$. The last variational equation ${\partial L_{\mathrm {eff}}}/{\partial \mathcal{N}}=0$ yields
\begin{eqnarray}
 \mu = \frac{1}{4w^2}+\sum_{i=1}^2\frac{A_i}{2}\left[1-\exp\left(-\alpha^2_iw^2\right)\right]+\frac{\gamma_0}{w\sqrt{2\pi}} +\sqrt{\frac{2}{\pi}}\frac{\gamma_1}{w\sqrt{w^2+4}} \nonumber \\
\hspace*{2cm} +\frac{\eta_0}{\pi \sqrt{3}w^2}+\frac{\eta_1}{3 \pi w^2}\frac{2 \sqrt{2}}{\sqrt{w^2+6}},
\label{eq:chem}
\end{eqnarray}
which defines the chemical potential.

\section{Numerical study}

To perform numerical simulation, we have employed split-step Crank-Nicolson (SSCN) method \cite{Muruganandam2009}. We have used both real and imaginary time propagation using adequately small space and time steps to obtain converged results. In practice, we have taken space and time steps as $0.0025$ and $0.00005$ respectively. For checking the consistency of the numerical calculation, we have compared our real-time propagation results with that of imaginary-time propagation and verified that the two are in good agreement with each other. The accuracy of our numerical program is also checked by varying space and time steps.

We have taken the strength ratio of the bi-chromatic optic lattice potential as $s_2/s_1 = 1$ throughout our investigation and the corresponding wavelengths are $\lambda_1 = 5$ and $\lambda_2 = 0.864\lambda1$~\cite{Roati2008}.

\subsection{Impact of spatially inhomogeneous binary interaction $(k(x)= 0)$.}

\begin{figure}[!ht]
\begin{center}
\includegraphics[width=\columnwidth, clip]{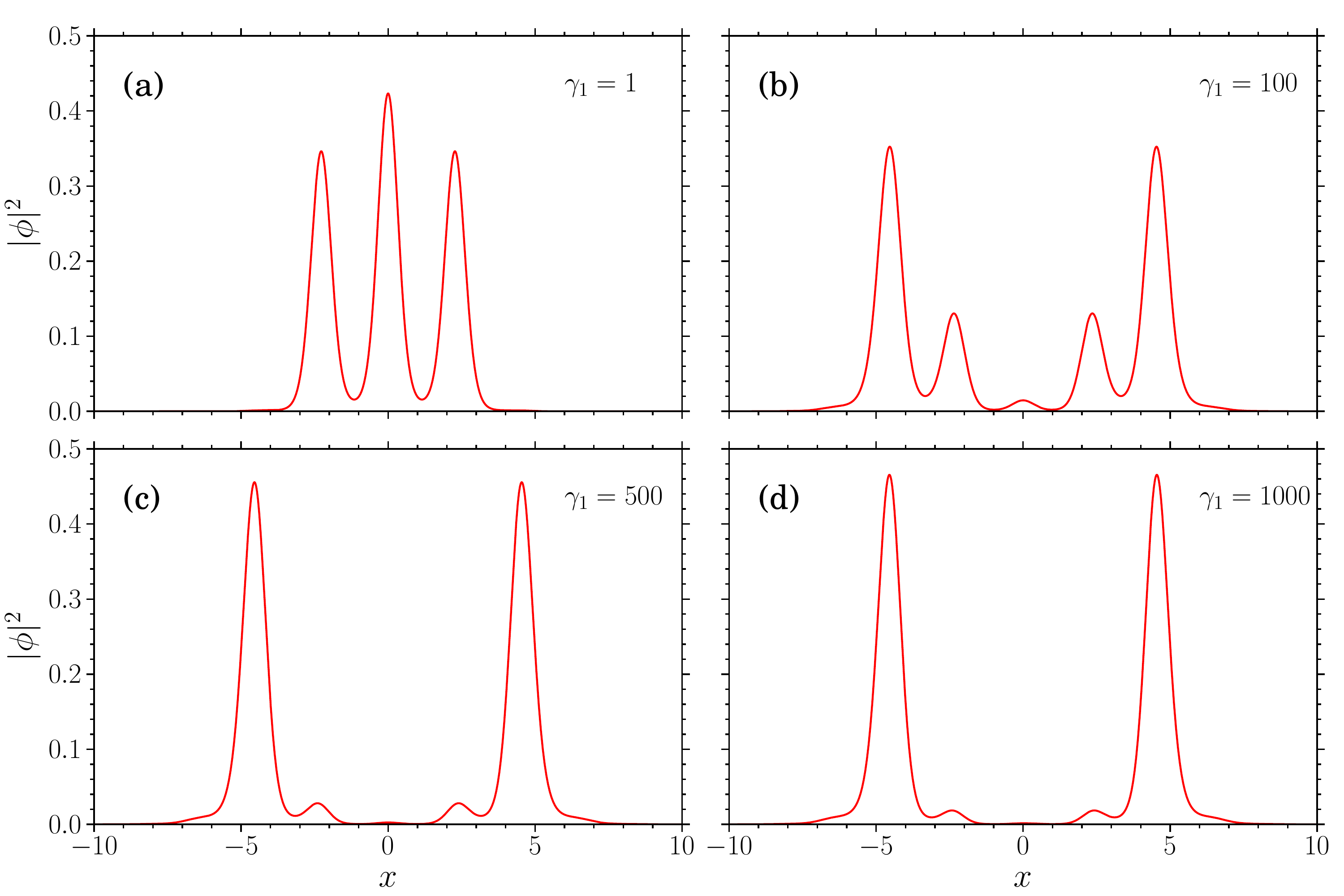}
\end{center}
\caption{Density profiles for repulsive binary spatially inhomogeneous strengths: (a) $\gamma_1= 1$, (b)  $\gamma_1=100$, (c) $\gamma_1=500$, (d) $\gamma_1=1000$ with the potential (\ref{pot:sin}).}
\label{fig2}
\end{figure}

\begin{figure}[!ht]
\begin{center}
\includegraphics[width=\columnwidth,clip]{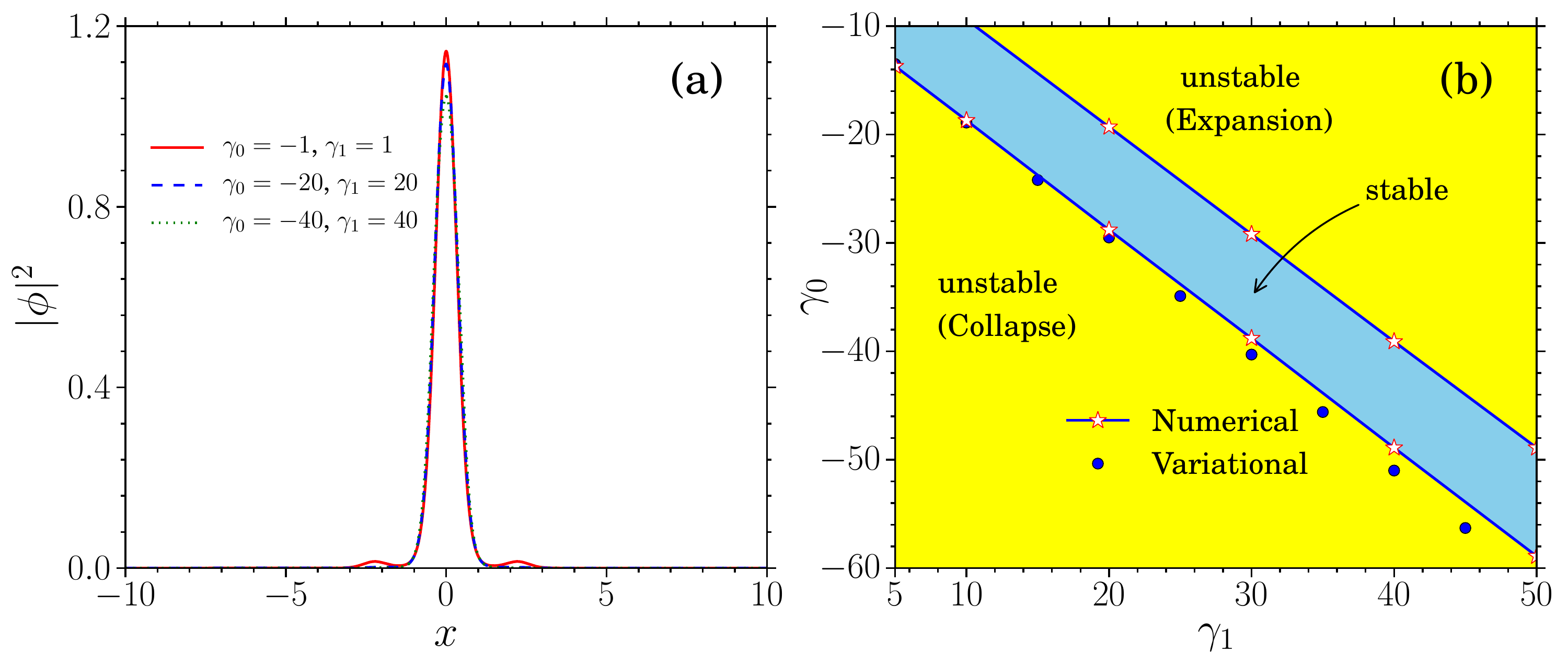}
\end{center}
\caption{(a) Density profiles for the potential (\ref{pot:sin}) and (b) phase plot showing regions of stability for repulsive binary spatially inhomogeneous interaction against constant binary attractive interaction.}
\label{fig3}
\end{figure}

\begin{figure}[!ht]
\begin{center}
\includegraphics[width=0.59\columnwidth,clip,trim=20 0 2 20]{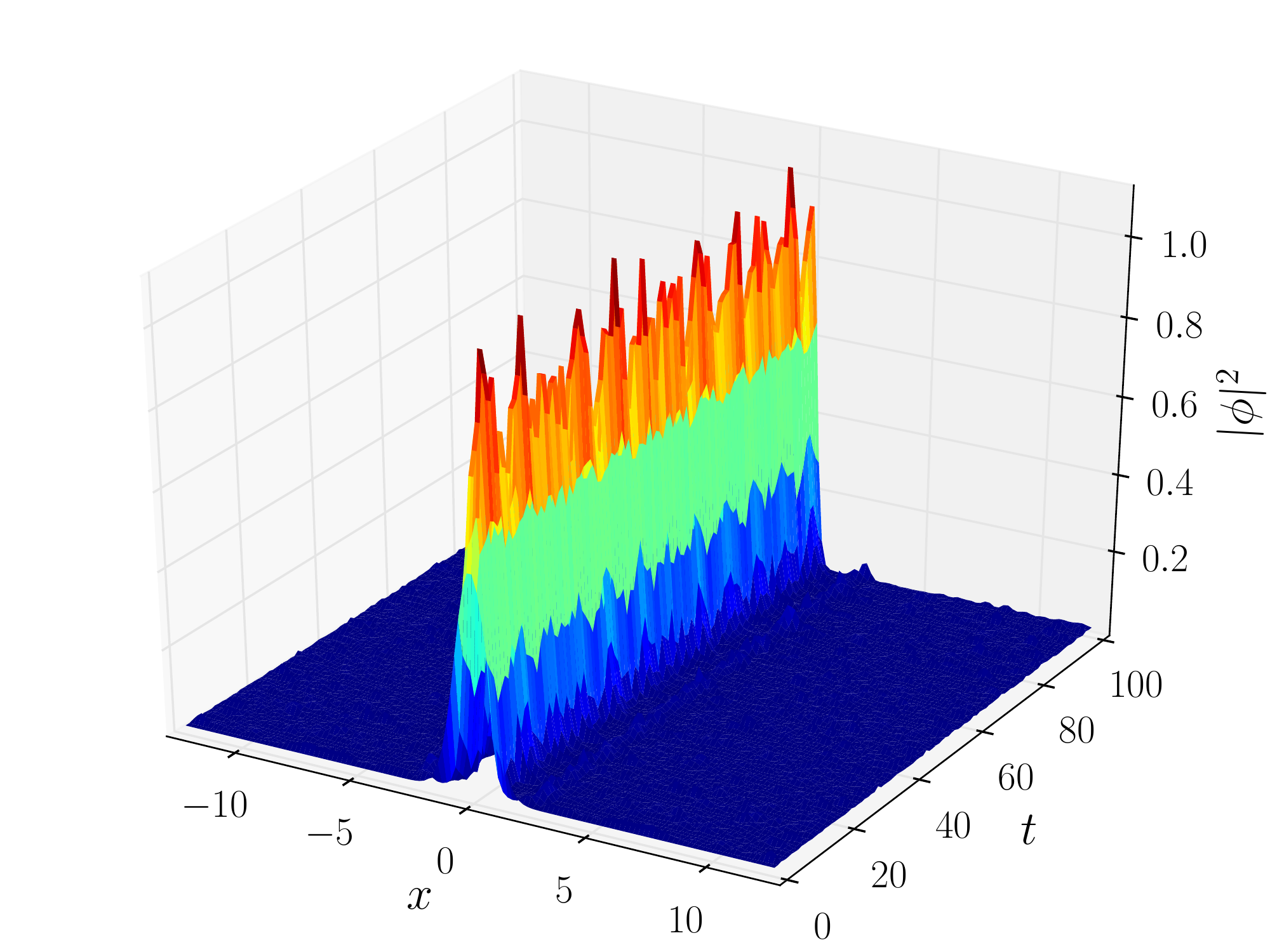}
\end{center}
\caption{The density profiles for $\gamma_0= -15$, $\gamma_1= 20$ for the potential (\ref{pot:sin}) as a function of time.}
\label{fig4}
\end{figure}

\begin{figure}[!ht]
\begin{center}
\includegraphics[width=0.98\columnwidth]{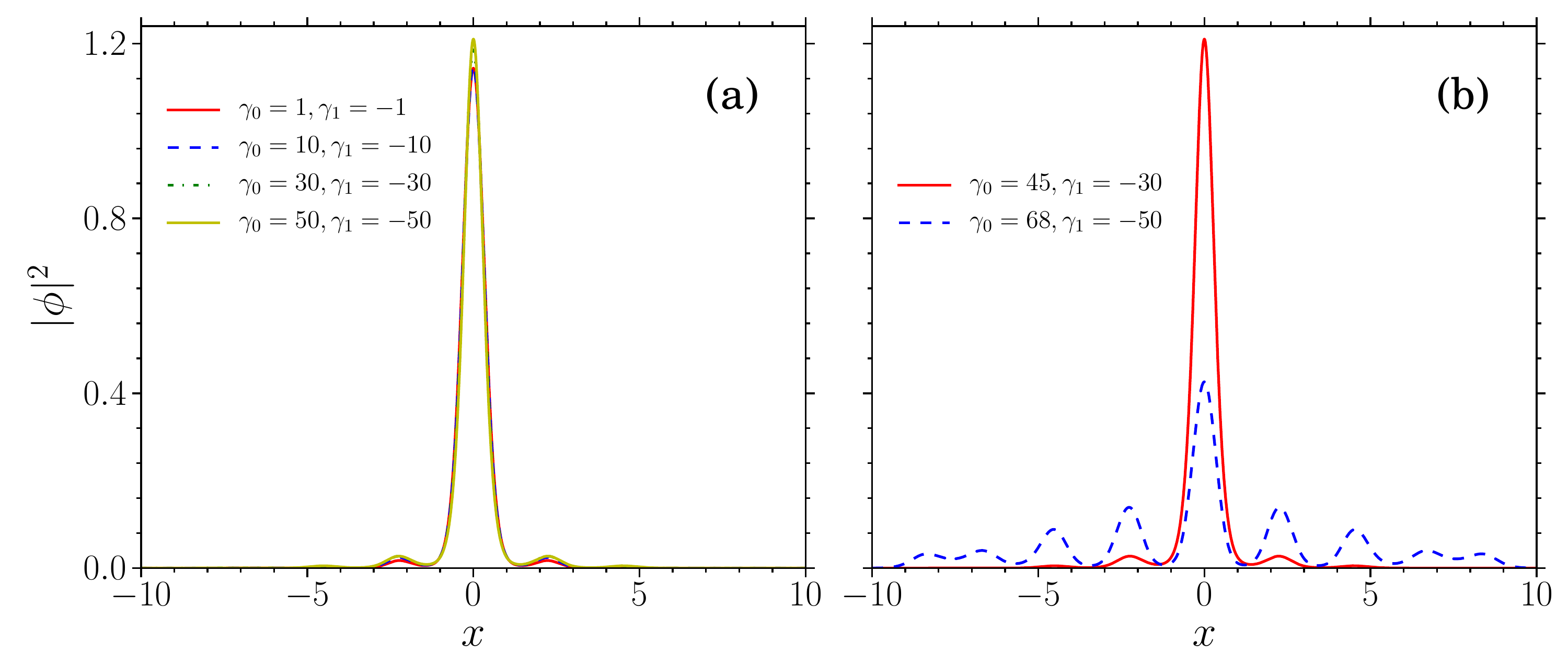}
\includegraphics[width=0.49\linewidth]{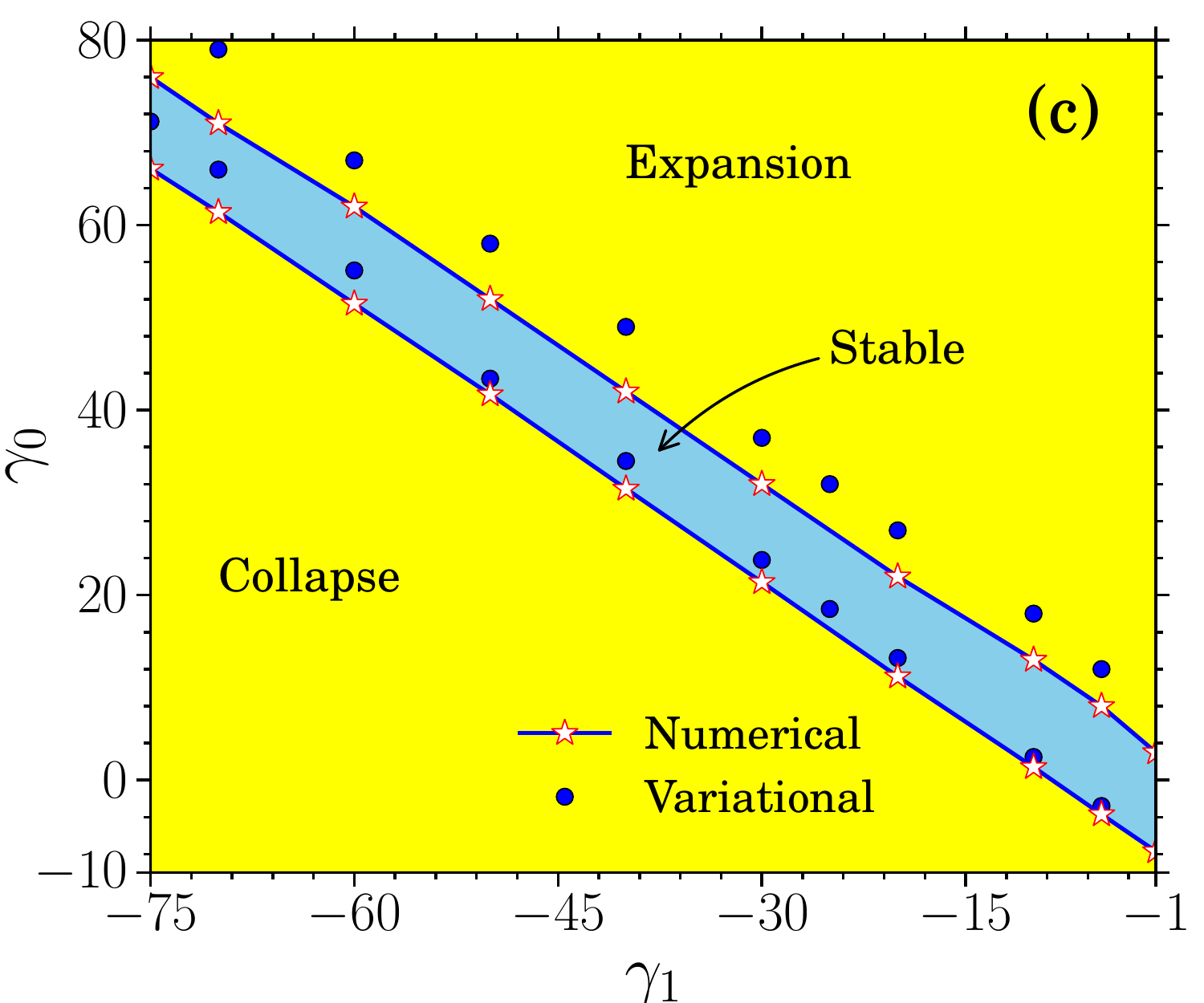}
\end{center}
%\begin{center}
%\includegraphics[width=0.49\linewidth]{npd0102}
%\end{center}
\caption{Plot of the density profiles for (a) stable and (b) expansion region for the potential (\ref{pot:sin}) and (c) the phase plot for attractive binary spatially inhomogeneous interaction against constant repulsive binary interaction.}
\label{fig5}
\end{figure}

The localization of BECs with a constant repulsive binary interaction has already been studied in  \cite{Adhikari2009}. So, we now introduce the  repulsive spatially inhomogeneous two body interaction ($\gamma_1$). The density profiles for various values of repulsive spatially inhomogeneous two body interactions are shown in figures \ref{fig2}(a)-(d). When we increase the binary repulsive spatially inhomogeneous nonlinearity,  instability sets in the condensates.
However, it should be mentioned that the amplitude of the central localized density can be controlled by adding an almost equal amount of constant attractive binary strength $(\gamma_0)$. The corresponding density profile is shown in figure \ref{fig3}(a) while the phase plot in figure \ref{fig3}(b) displays the range of constant attractive binary strengths which could be reinforced with repulsive spatially inhomogeneous binary strengths to sustain central localized hump. The density profile shown in figure \ref{fig3}(a) represents the solitary mode.
Figure \ref{fig4} shows the density profile for $\gamma_0= -15$ and $\gamma_1= 20$ for various time intervals corresponding to the domain above the stable region as shown in figure \ref{fig3}(b). Eventhough the density profile in figure \ref{fig4} seems to indicate the presence of a dynamically stable state, the fluctuations around the boundaries contribute to the increase in the rms size thereby leading to the expansion of BECs. In the region below the stable domain, the central hump density increases with $\gamma_0$ and blows up at a critical value of $\gamma_0$. %\st{From the above, one also observes that the constant binary attractive interaction $\gamma_0$ prevails over the repulsive spatially inhomogeneous interaction $\gamma_1$.}
\paragraph{•}
One can also sustain the central localized density by adding almost equal amounts of constant repulsive binary strength with the attractive spatially inhomogeneous binary interaction as shown in figure \ref{fig5}(a). From figure \ref{fig5}(a), we understand that the central localized stable density profile corresponds to the solitary mode while the density begins to occupy multiple sites on either sides of the center for increasing constant repulsive binary nonlinearity as shown in figure \ref{fig5}(b). The numerical domain of values for constant repulsive binary strength against the attractive spatially inhomogeneous binary strength is shown in figure \ref{fig5}(c).
%\newpage

\subsection{Impact of both spatially inhomogeneous binary and spatially inhomogeneous three body interactions}

\begin{figure}[!ht]
\begin{center}
\includegraphics[width=\columnwidth,clip]{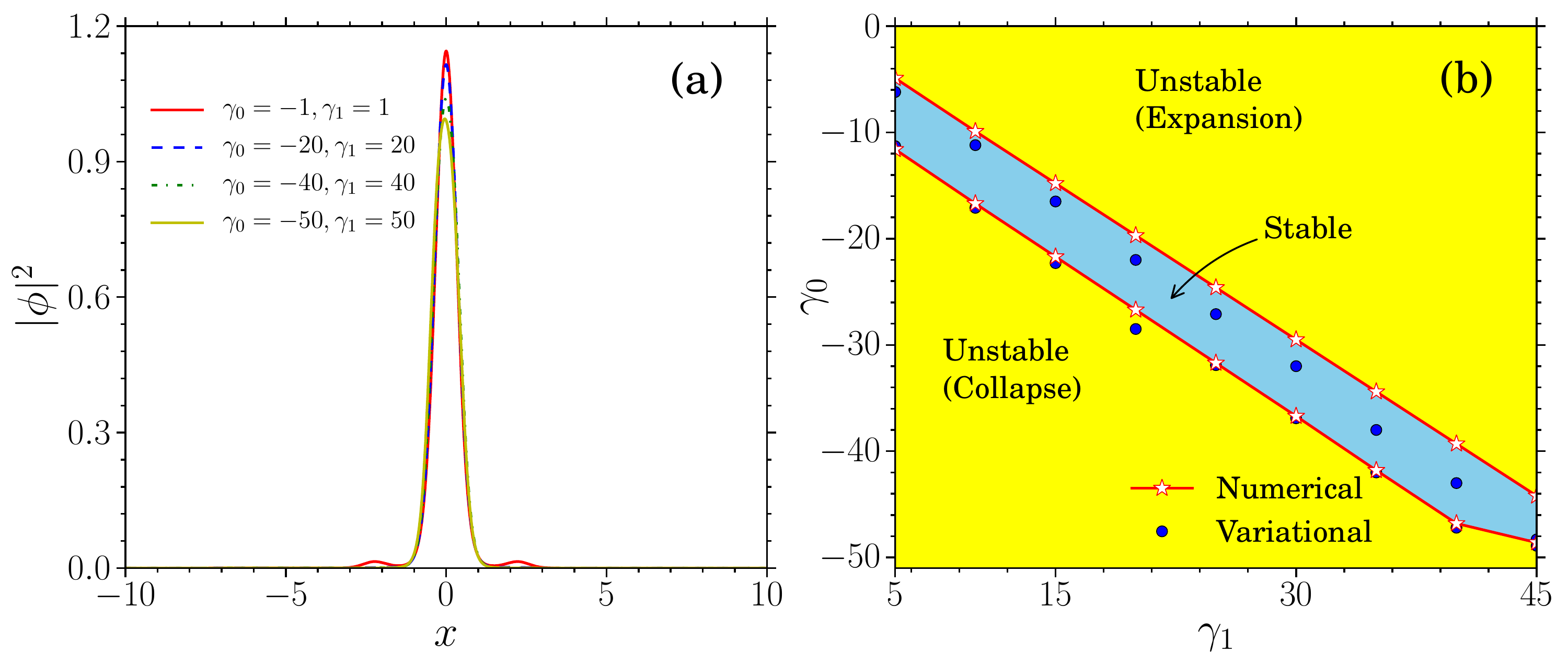}
\end{center}
\caption{{(a) Plot of the density profiles corresponding to the potential (\ref{pot:sin}) for different $\gamma_0$ and $\gamma_1$, and (b) the phase plot for repulsive spatially inhomogeneous binary and repulsive spatially inhomogeneous three body interactions against constant attractive binary and constant attractive three body interactions.}}
\label{fig6}
\end{figure}

\begin{figure}[!ht]
\begin{center}
\includegraphics[width= \columnwidth,clip]{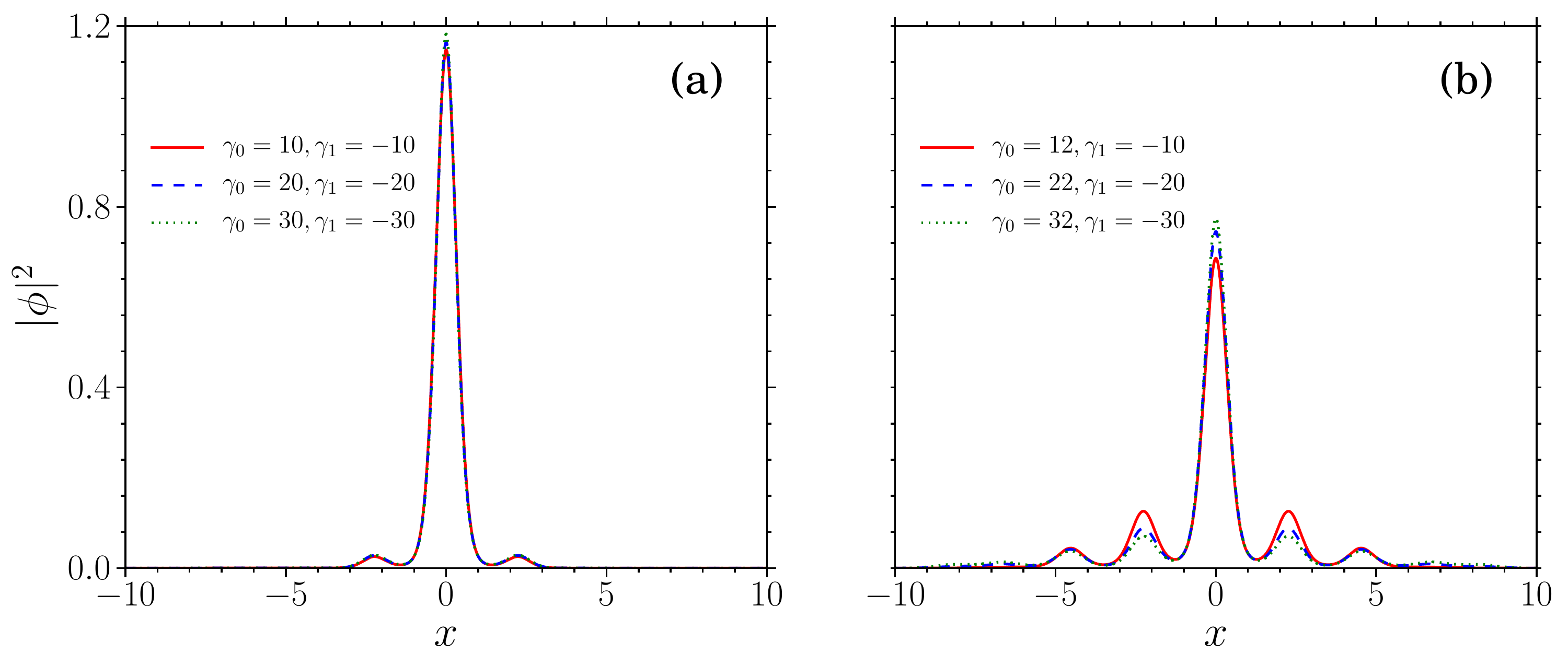}
\end{center}
\caption{Plot of the density profiles in the (a) stable and (b) expansion regions for the potential (\ref{pot:sin}).}
\label{fig7}
\end{figure}

\begin{figure}[!ht]
\begin{center}
\includegraphics[width=0.49\linewidth]{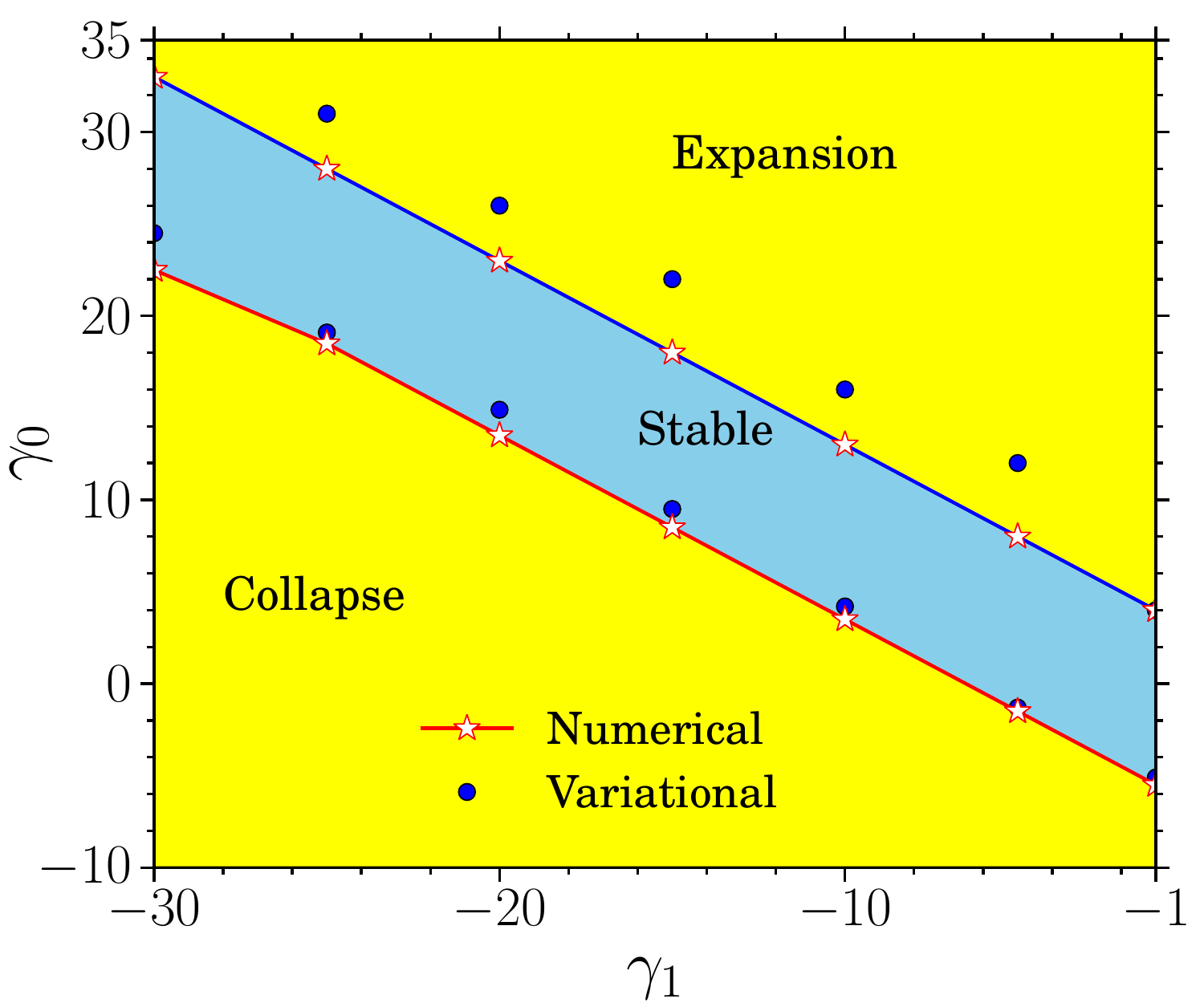}
\end{center}
\caption{The phase plot for attractive spatially inhomogeneous binary and three body interactions against constant repulsive binary and three body interactions.}
\label{fig8}
\end{figure}

We now introduce the three body interaction (both constant and spatially inhomogeneous) in the presence of both constant and spatially inhomogeneous two body interaction on the condensates. We now choose a weak three body interaction $k(x)= 0.1 g(x)$ in phase with the binary interaction. Figures \ref{fig6}(a) and \ref{fig6}(b) show the density and phase plots for both repulsive spatially inhomogeneous binary and repulsive spatially inhomogeneous three body interaction against constant attractive binary and constant attractive three body interaction.
%\newpage
The density plot for the stable region in figure \ref{fig6}(b) which is shown in figure \ref{fig6}(a) represents the solitary mode. The condensate density in figure \ref{fig6}(a) is quite identical to figure \ref{fig3}(a). This means that in the stable region, the condensate density is not at all affected by the weak three body interaction. In the domain above the stable region shown in figure \ref{fig6}(b), the condensates expand quickly as compared to that of BECs with $k(x)=0$ similar to what is transpiring corresponding to the domain above the stable region shown in figure \ref{fig3}(b). Reversing the sign of $\gamma_0$ and $\gamma_1$, one obtains the central localized solitary mode as shown in figure \ref{fig7}(a) when there is perfect balance between $\gamma_0$ and $\gamma_1$. Any imbalance between $\gamma_0$ and $\gamma_1$ may either lead to the multiple occupation of the condensates at various sites on either sides shown in figure \ref{fig7}(b) corresponding to the domain above the stable region of figure \ref{fig8} or the collapse of the condensates corresponding to the domain below the stable region shown in figure \ref{fig8}. Comparing the phase plots shown in figures 2(b) and 4(c) with figures 5(b) and 7, respectively, we observe that the condensates rapidly move into the unstable region under the combined impact of spatially inhomogeneous binary and spatially inhomogeneous three body interaction.
\newpage

\section{Conclusion}

In this paper, we have investigated the combined impact of both spatially inhomogeneous binary and spatially inhomogeneous three body interactions on BECs in a bichromatic optical lattice. Our results show that the condensates which becomes unstable after the introduction of repulsive spatially inhomogeneous binary interaction can be stabilized by the addition of constant attractive binary strength in almost equal amounts. If the balance between repulsive spatially inhomogeneous binary interaction and constant attractive interaction is disturbed, the condensates become unstable. If the sign of the interaction is reversed, the condensate density is also found to be stable for almost equal strengths. Any imbalance between attractive spatially inhomogeneous interaction and constant repulsive interaction either results in the collapse of BECs or in the occupation of the condensates at multiple sites on either sides.  When we introduce a weak three body interaction in phase with the binary interaction (repulsive spatially inhomogeneous binary and repulsive spatially inhomogeneous three body against constant attractive binary and constant attractive three body), the extent of instability increases. Reversing the sign of spatially inhomogeneous and constant interaction enhances the stability of BECs.

\ack

JBS wishes to acknowledge financial assistance received from Department of Science and Technology (Ref. No. SR/S2/HEP-26/2012).  The work of RR forms a part of University Grants Commission (Ref. No. UGC-40-420/2011 (SR)), Department of Atomic Energy - National Board of Higher Mathematics (Ref. No. DAE-NBHM-2/ 48(1)2010/NBHM-RD II/4524) and Department of Science and Technology (Ref. No. SR/S2/HEP-26/2012), Government of India sponsored projects. The work of PM forms a part of Department of Science and Technology (Ref. No. SR/S2/H EP-03/2009) and Council of Scientific and Industrial Research (Ref. No. 03(1186)/10/EMR-II), Government of India funded research projects. Authors thank the anonymous referees for their constructive comments and suggestions to improve the manuscript.

 \end{document}